\documentclass[aps,prb,twocolumn,superscriptaddress,floatfix,letter,longbibliography]{revtex4-2}

\usepackage{epsfig,amsmath,amssymb,color,amsfonts,physics,microtype}
\usepackage[bookmarks=true,colorlinks,citecolor=red]{hyperref}
\usepackage{dsfont}
\usepackage{wrapfig}
\usepackage{tikz}
\usepackage{algorithm}
\usepackage{algpseudocode}

\usepackage{tikz-network}
\usepackage{bbold}

\usetikzlibrary {shapes.geometric,positioning}
 \tikzset{projector/.style={isosceles triangle, draw, inner sep=0pt,
     anchor=south, shape border rotate=90, isosceles triangle stretches}}

\DeclareMathOperator*{\argmax}{argmax}
\newcommand{\daga}{^\dag}
\newcommand{\ssigma}{\boldsymbol{\sigma}}

\begin{document}

\title{Estimating Non-Stabilizerness Dynamics Without Simulating It}

\author{Alessio Paviglianiti}
\affiliation{International School for Advanced Studies (SISSA), via Bonomea 265, 34136 Trieste, Italy}
\author{Guglielmo Lami}
\affiliation{International School for Advanced Studies (SISSA), via Bonomea 265, 34136 Trieste, Italy}
\affiliation{Laboratoire de Physique Théorique et Modélisation, CY Cergy Paris Université, CNRS, F-95302 Cergy-Pontoise, France}
\author{Mario Collura}
\affiliation{International School for Advanced Studies (SISSA), via Bonomea 265, 34136 Trieste, Italy}
\affiliation{INFN Sezione di Trieste, 34136 Trieste, Italy}
\author{Alessandro Silva}
\affiliation{International School for Advanced Studies (SISSA), via Bonomea 265, 34136 Trieste, Italy}

\begin{abstract}
We introduce the Iterative Clifford Circuit Renormalization (ICCR), a novel technique designed to efficiently handle the dynamics of non-stabilizerness (a.k.a. quantum magic) in generic quantum circuits. ICCR iteratively adjusts the starting circuit, transforming it into a Clifford circuit where all elements that can alter the non-stabilizerness, such as measurements or $T$ gates, have been removed. In the process the initial state is renormalized in such a way that the new circuit outputs the same final state as the original one. This approach embeds the complex dynamics of non-stabilizerness in the flow of an effective initial state, enabling its efficient evaluation while avoiding the need for direct and computationally expensive simulation of the original circuit. The initial state renormalization can be computed explicitly using a matrix-product state approximation that can be systematically improved. We implement the ICCR algorithm to evaluate the non-stabilizerness dynamics for systems of size up to $N=1000$. We validate our method by comparing it to tensor networks simulations. Finally, we employ the ICCR technique to study a magic purification circuit, where a measurement-induced transition is observed.
\end{abstract}

\maketitle

\section{Introduction}\label{s:introduction}
Investigating the physics of quantum many-body systems using classical simulation is a notoriously challenging in both condensed matter and statistical physics~\cite{Feynman1982}. The amount of resources needed to represent a many-body quantum state grows exponentially with the number of degrees of freedom involved, thus limiting numerical investigations to small system sizes~\cite{Kohn_1999}. Nevertheless, there exist particular classes of quantum states, such as lowly entangled~\cite{Silvi_2019,Biamonte_2020,Vidal_2004,Hastings_2007,Schollwock_2011}, Gaussian~\cite{Terhal_2002,Jozsa_2008,Surace_2022}, and stabilizer states~\cite{Gottesman_1997,Gottesman_1998_1,Gottesman_1998_2,Aaronson_2004}, that are simple enough to allow for their efficient simulation on classical devices, yet complex enough to exhibit genuine many-body phenomena such as quantum criticality and scrambling. The existence of such special states raised the broad problem of classifying and quantifying the complexity of a quantum state, which in parallel has become a useful way to characterize and distinguish different many-body phases.

Currently there is no unique definition of state complexity, as states have many ways to be non-trivial. For $N$-qubit systems, a possible characterization is given by the so-called non-stabilizerness, a.k.a. \textit{quantum magic}~\cite{Bravyi_2005,Seddon_2021,Veitch_2014,Howard_2017,Winter_2022}, which provides a measure of how far a given state is from the class of easy-to-simulate stabilizer states and directly bounds the computational efficiency of its numerical simulation. Multiple measures of magic have been proposed and used to investigate circuit~\cite{Oliviero_2022,Niroula_2023,Turkeshi_2023,Fux_2023,Bejan_2023,Mello_2024} and many-body problems~\cite{Lami_2023,Lami_2024,Fux_2023,Tarabunga_2023}. Despite this progress, the investigation of many-body non-stabilizerness is computationally challenging: not only non-stabilizer states are hard to simulate in the first place, but also the known magic measures are exponentially expensive to evaluate~\cite{Veitch_2014,Leone_2022,Haug_2023_3}. While this problem can be mitigated using efficient non-stabilizerness estimation algorithms for Matrix Product States (MPS)~\cite{Lami_2023,Lami_2024,Haug_2023_1,Tarabunga_2023,Tarabunga_2024}, these limitations hinder the investigation of regimes inaccessible to tensor network methods, such as very large system sizes, higher dimensionality, or highly entangled states.

In this work, we propose a novel technique, which we name ``Iterative Clifford circuit renormalization" (ICCR), to address magic efficiently bypassing the aforementioned limitations. The method can be applied to states obtained from any given initial state through Clifford circuits doped with non-stabilizer gates and possibly measurements, in principle with arbitrary dimensionality. The main advantage of our approach is that we avoid entirely to evaluate the evolution generated by the circuit by iteratively manipulating its geometry and renormalizing the initial state. In practice, we use a simple yet effective approximation method to follow the renormalization flow of the initial state, finding an optimal description as a matrix-product state (MPS). This approach allows for a systematic improvement of the approximation by simply increasing the bond dimension $\chi$ of the MPS.

The ICCR algorithm combined with this approximation enables the exploration of system sizes up to $N=1000$. We demonstrate the validity of this technique by studying the problem of a 1D monitored Clifford circuit where the initial state has extensive non-stabilizerness, and we highlight a measurement-induced transition~\cite{Skinner_2019,Gullans_2020,Turkeshi_2020,Nahum_2021,Zabalo_2020,Paviglianiti_2023,Sierant_2022} in the relaxation dynamics of magic measures tuned by the measurement rate. To validate our method, we investigate the convergence of the non-stabilizerness estimates with $\chi$, obtaining that the overall error scales as $\sim \chi^{-1}$. In addition, we compare our results to exact tensor network simulations, showing good agreement. Finally, our numerics suggest that even when the magic estimates are not accurate they still provide lower bounds to the exact ones.

The rest of this paper is structured as follows. In Sec.~\ref{s:magic} we introduce the notion of non-stabilizerness, as well as its measures. Then, Sec.~\ref{s:algorithm} presents the ICCR algorithm to compute these measures efficiently, providing a step-by-step derivation and a discussion on its performance. We report an example of application of this formalism in Sec.~\ref{s:numerics}, which introduces the model we consider, our results, an an analysis of the error of the approximation, and the benchmark of our method with tensor network simulations. Finally, we summarize our findings in Sec.~\ref{s:conclusions}.

\section{Non-stabilizerness Measures}\label{s:magic}
In this section we briefly review the concept of magic and its measures. First, we define stabilizer states and their main features. We then introduce the non-stabilizerness measures that are relevant for our investigation, namely the stabilizer nullity and the Stabilizer Renyi Entropies (SREs)~\cite{Leone_2022,Lami_2023,Oliviero_2022_1,Haug_2023_2,Tirrito_2023,Rattacaso_2023}, commenting on their properties. For convenience, throughout the manuscript we denote the eigenstates of the Pauli operator $\hat{Z}$ by $\ket{+}$ and $\ket{-}$ (instead of $\ket{0}$ and $\ket{1}$, respectively).

Consider a system of $N$ qubits, and define the $N$-qubit Pauli group $\mathcal{P}_N = \{\pm 1,\pm i\}\{\hat{\mathds{1}},\hat{X},\hat{Y},\hat{Z}\}^{\otimes N}$. The class of stabilizer states is the set of states $\ket{\Psi}$ for which there exist $N$ commuting and independent Pauli strings $\hat{g}_i\in \mathcal{P}_N$, $i=1,\dots,N$, such that $\hat{g}_i\ket{\Psi}=+\ket{\Psi}$. The operators $\hat{g}_i$ are said to stabilize the state, and they generate the stabilizer group of 
$\ket{\Psi}$. The knowledge of the $N$ generators completely defines the state. There exists a set of unitary gates, called the Clifford group $\mathcal{C}_N$, that maps stabilizer states into other stabilizer states (and Pauli strings into other Pauli strings). This group is generated by only three elementary gates, namely the Hadamard gate $\hat{H}$, the phase gate $\hat{S}$, and the CNOT gate $\hat{\mathrm{CX}}$.

By applying a non-Clifford gate $\hat{U}\notin\mathcal{C}_N$ to a stabilizer state in general some or all the generators $\hat{g}_i$ will be mapped to $\hat{g}'_i = \hat{U}\hat{g}_i\hat{U}\daga\notin\mathcal{P}_N$, thus making the state no longer a stabilizer. The amount of magic of a state can be quantified through different measures. A good measure $\mathcal{M}(\ket{\Psi})$ should fulfill some properties. First, $\mathcal{M}(\ket{\Psi_\text{stab}})=0$ if and only if $\ket{\Psi_\text{stab}}$ is a stabilizer state must have $\mathcal{M}(\ket{\Psi_\text{stab}})=0$, and $\mathcal{M}(\ket{\Psi})> 0$ otherwise. Second, it must not change under the application of any Clifford operation $\hat{U}_\text{Cliff}\in\mathcal{C}_N$, namely, $\mathcal{M}(\hat{U}_\text{Cliff}\ket{\Psi})=\mathcal{M}(\ket{\Psi})$. Finally, it must be additive for separable states, such that $\mathcal{M}(\ket{\Psi_1}\otimes\ket{\Psi_2}) = \mathcal{M}(\ket{\Psi_1})+\mathcal{M}(\ket{\Psi_2})$.

\begin{figure}[t!]
    \centering
    \includegraphics[width=0.65\columnwidth]{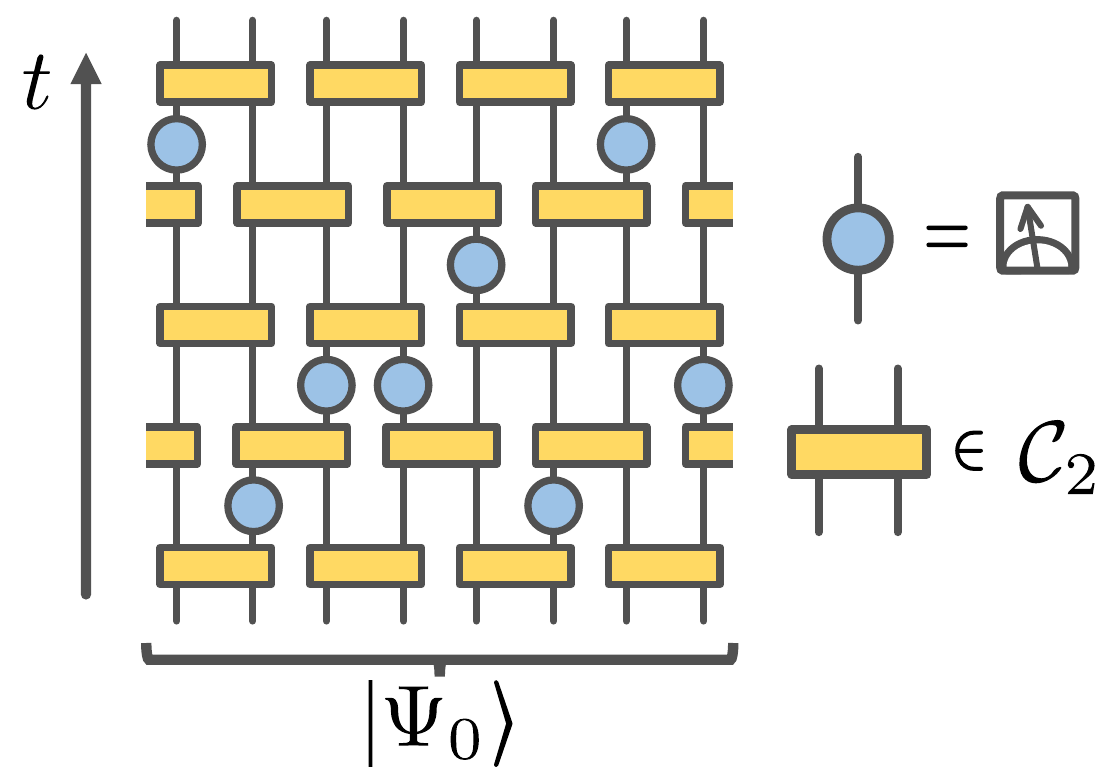}
    \caption{Example of Clifford circuit with measurements. Measurements of Pauli strings are represented by blue rounded boxes, unitary operators by yellow squared boxes.}
    \label{f:circuit}
\end{figure}
In our analysis we consider different non-stabilizerness measures. A first measure is the stabilizer nullity $\nu$, which is defined in terms of the stabilizer rank $r$. The stabilizer rank of a state $\ket{\Psi}$ is defined as the maximum integer $R$ for which it is possible to write
\begin{equation}\label{stab_rank}
    \ket{\Psi} = \hat{U}_\text{Cliff}\left(\ket{+}^{\otimes R} \ket{\Phi^{(N-R)}}\right),
\end{equation}
where $\hat{U}_\text{Cliff}$ is a suitable Clifford operator, and $\ket{\Phi^{(N-R)}}$ is a state of $N-R$ qubits. The rank $r$ coincides with the number of generators $\hat{g}_i\in\mathcal{P}_N$ of the stabilizer group of $\ket{\Psi}$, and it quantifies how many single-qubit stabilizer states it is possible to distill out of $\ket{\Psi}$ using only Clifford operations. The stabilizer nullity is then defined as $\mathcal{\nu}=N-r$. We also consider the SREs~\cite{Leone_2022} given by
\begin{equation}\label{sres}
    M_n = \frac{1}{1-n}\log_2\left(\sum_{\hat{P}\in\Tilde{\mathcal{P}}_N}\frac{1}{2^N}\bra{\Psi}\hat{P}\ket{\Psi}^{2n}\right),
\end{equation}
where $\Tilde{\mathcal{P}}_N=\{\hat{\mathds{1}},\hat{X},\hat{Y},\hat{Z}\}^{\otimes N}$ is the set of Pauli strings taken without phases. Differently from the nullity, the SREs assume continuous values and they are proper monotones for $n\geq 2$~\cite{Haug_2023_2,leone2024stabilizer}. The nullity is formally given by $\nu = \lim_{n\to\infty} (n-1) M_n$, and can be proven to set an upper bound to $M_n$~\cite{Tarabunga_2024}.

\section{Iterative Clifford circuit renormalization}\label{s:algorithm}
\begin{figure}[t!]
    \centering
    \includegraphics[width=0.4\columnwidth]{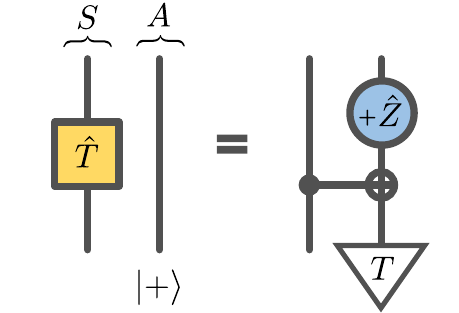}
    \caption{Graphic representation of the T gadget replacement. The blue circle represents a projective measurement of $\hat{Z}$ with positive outcome, as in Fig.~\ref{f:circuit}.}
    \label{f:T_gadget}
\end{figure}

In this section we describe the ICCR technique we developed to study the magic dynamics. This method can be applied to a wide class of perturbed Clifford circuits where the magic is either introduced in the initial state or directly injected through non-Clifford $\hat T = {\rm diag}\{1,e^{i\pi/4}\}$ gates. We treat both possibilities in a unified framework, reducing both cases to a non-stabilizer initial state $\ket{\Psi_0}$ evolved with only Clifford operators and computational-basis measurements, as represented schematically in Fig.~\ref{f:circuit}. This is achieved by making use of the so-called $T$ gadget~\cite{Bravyi_2016,Zhou_2000}, which allows us to translate a $\hat T$ gate into an equivalent circuit involving a projective measurement and an ancilla qubit prepared in a resource state. In detail, suppose we want to apply a $\hat T$ gate to a generic state $\ket{\Psi_S}$ of the system $S$. We introduce an ancilla qubit $A$ prepared in the resource state $\ket{T_A} = (\ket{+}+e^{i\pi/4}\ket{-})/\sqrt{2}$. The following identity can be easily verified 
\begin{equation}\label{T_gadget}
    (\hat{T}_S \ket{\Psi_S})\ket{+_A} = \frac{1}{\sqrt{2}}\frac{\hat{\mathds{1}}+\hat{Z}_A}{2}\hat{\mathrm{CX}}_{S\to A} \big(\ket{\Psi_S}\ket{T_A}\big),
\end{equation} 
The T gadget is graphically represented in Fig.~\ref{f:T_gadget}. Therefore, it is always possible to replace $\hat T$ gates by supplementing the system with ancillae and performing Clifford operations and projective measurements.

\begin{figure*}[t!]
    \centering
    \includegraphics[width=\textwidth]{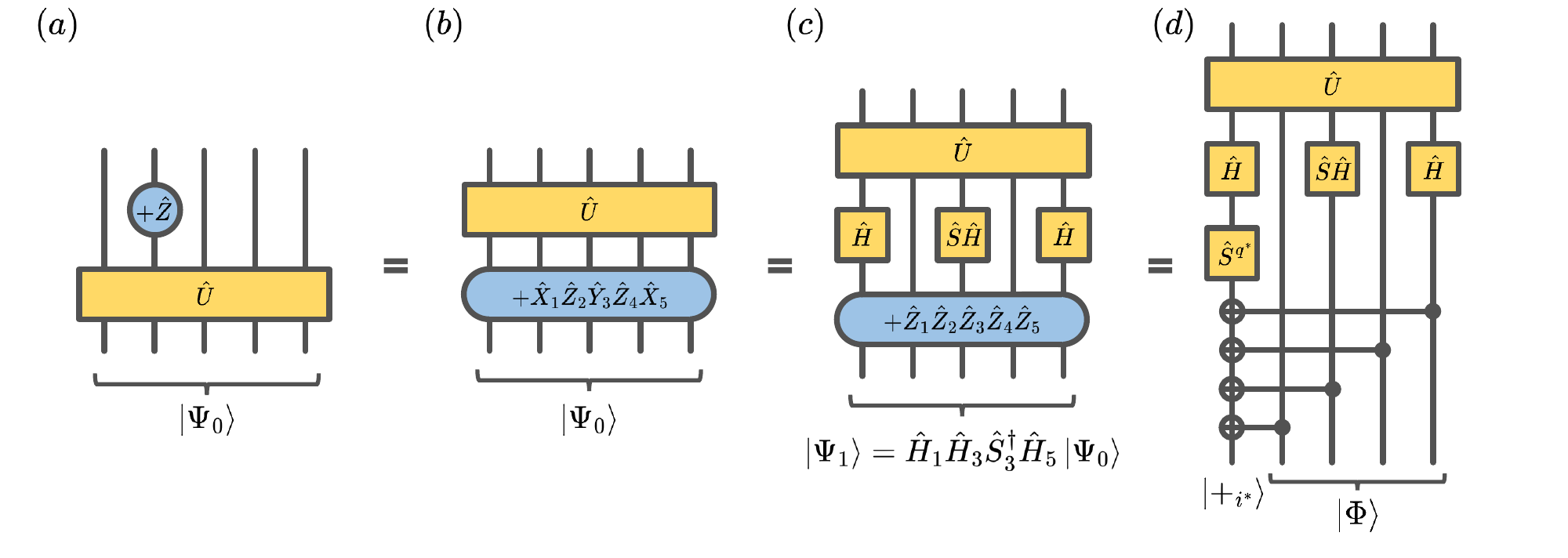}
    \caption{Schematic representation of the ICCR algorithm. Projectors on Pauli strings are represented by blue rounded boxes, unitary operators by yellow squared boxes. (a) The original circuit involves a unitary Clifford gate followed by a local projector, in this example corresponding to $\frac{(\hat{\mathds{1}}\pm\hat{Z}_2)}{2}$. (b) The projector is moved through the Clifford gate resulting in a measurement of an extended Pauli string. (c) Single-qubit gates are applied to rotate the new Pauli string into one involving only $\hat{Z}$. (d) The projector is replaced by a cascade of CNOT and single-qubit gates, and the initial state is updated.}
    \label{f:algo_scheme}
\end{figure*}

Below we provide a description of the ICCR method, schematized pictorially in Fig.~\ref{f:algo_scheme}. We start by providing a summary of the technique aimed at clarifying its conceptual procedure and physical intuition. We then explain the details needed for a practical implementation of the method. Finally, we comment on its computational cost and performance.

\subsection{ICCR in a nutshell}
The concept underlying the ICCR algorithm is the following. Given a non-stabilizer initial state $\ket{\Psi_0}$ evolved with Clifford unitary gates and projective measurements, we want to obtain a renormalized initial state $\ket{\Psi'}$ and a measurement-free Clifford circuit $\hat{U'}$ such that $\hat{U}'\ket{\Psi'}$ outputs the same state as the original circuit. If successful, this approach enables direct evaluation of magic measures on $\ket{\Psi'}$, bypassing the need of simulating the entire time evolution. The ICCR algorithm performs the task iteratively, i.e.\ by removing one measurement at a time from the original circuit until none is left. 

An iteration of the algorithm proceeds as follows. We consider the initial state $\ket{\Psi_0}$ and the portion of the unitary Clifford circuit $\hat{U}_0$ up to the first projective measurement (as in Fig.~\ref{f:algo_scheme}a), which we assume to be of $\hat{Z}_j$ at a certain position $j$. We swap measurement and unitary gate to obtain an equivalent circuit with a measurement acting directly on $\ket{\Psi_0}$. This transformation can always be accomplished efficiently, since $\hat{U}_0$ is Clifford. The result will yield a projective measurement of a new Pauli string $\hat{P}$ spread across multiple qubits. 

The key idea of the ICCR method is to replace the new projector with an appropriate sequence of Clifford unitary gates $\hat{U}'$ acting on a renormalized initial state $\ket{\Psi'}$. In other words, denoting by $\hat{\Pi}=(\hat{\mathds{1}}+s\hat{P})/2$ the projector associated to the measurement of $\hat{P}$ with outcome $s=\pm$, we want to rewrite
\begin{equation}
    \hat{\Pi}\ket{\Psi_0}=\mathcal{N}\hat{U}'\ket{\Psi'},
\end{equation}
where $\mathcal{N}$ is the norm of the left-hand side. A key observation to this is that, regardless of the outcome $s$, the projected state obeys a constraint that effectively removes one qubit degree of freedom. This request can be always mimicked by taking $\ket{\Psi'} = \ket{+}\ket{\Phi}$ and $\hat{U}'$ as a sequence of CNOT gates entangling $\ket{+}$ with the remaining degrees of freedom (see Fig.~\ref{f:algo_scheme}d). The price to pay is that $\ket{\Phi}$ is in general an unknown complex state. The key approximation would be to determine it variationally by minimizing the distance between $\ket{+}\ket{\Phi}$ and $(\hat{U}')\daga\hat{\Pi}\ket{\Psi_0}$.

\subsection{Description of the implementation}\label{ss:algorithm_description}
We now explain step by step how to implement the ICCR method. A schematic description can be found in Algorithm~\ref{alg:ICCR}. The basic iteration of the algorithm can be understood by considering an initial state $\ket{\Psi_0}$ with rank $r$ that is evolved by a Clifford unitary $\hat{U}_0$ followed by a projective measurement of the operator $\hat{Z}_j$, as represented in Fig.~\ref{f:algo_scheme}a. Taking into account the definition of the stabilizer rank (cf. Eq.~\eqref{stab_rank}), we can assume without any loss of generality that
\begin{equation}\label{psi0}
    \ket{\Psi_0} = \bigotimes_{i=1}^r\ket{\varphi_i}\ket{\Phi^{(N-r)}_\mathrm{non-stab}},
\end{equation}
i.e., the first $r$ spins occupy single-qubit stabilizer states. We would like the projector to act directly on the initial state, so we swap it with the unitary $\hat{U}_0$ by introducing the Pauli string
\begin{equation}\label{P_string}
    \hat{P} = \hat{U}_0\daga\hat{Z}_j\hat{U}_0,
\end{equation}
which allows us to rewrite
\begin{equation}\label{swapped_meas}
    \frac{(\hat{\mathds{1}}\pm\hat{Z}_j)}{2}\hat{U}_0\ket{\Psi_0} = \hat{U}_0\frac{(\hat{\mathds{1}}\pm\hat{P})}{2}\ket{\Psi_0},
\end{equation}
as shown in Fig.~\ref{f:algo_scheme}b.
\begin{algorithm}[H]
\caption{ICCR iteration} \label{alg:ICCR}
\textbf{Input:} the initial state $\ket{\Psi_0}=\bigotimes_{i=1}^r\ket{\varphi_i}\ket{\Phi^{(N-r)}_\text{non-stab}}$, a Clifford unitary $\hat{U}_0$ and the measured operator $\hat{Z}_j$.
   \begin{algorithmic}[1]
   \State Compute $\hat{P}=\hat{U}_0\daga\hat{Z}_j\hat{U}_0$.
   \State If possible, simplify $\hat{P}$ by checking its action on $\ket{\varphi_i}$.
   \If{$\hat{P}\ket{\Psi_0}=\pm\ket{\Psi_0}$}
   \State The measurement leaves the state unchanged, \textbf{break}.
   \EndIf
   \State Find Cliffords $\hat{q}_i$ such that $\left(\prod_i\hat{q}_i\right)\hat{P}\left(\prod_i\hat{q}_i\daga\right) = \prod_{i \in \mathcal{S}} \hat{Z}_i$
   \State Define $\ket{\Psi_1}=\left(\prod_i\hat{q}_i\right)\ket{\Psi_0}$ and $\hat{U}_1=\hat{U}_0\left(\prod_i\hat{q}_i\daga\right)$.
   \State If needed, pick a random outcome $s$ using the Born rule.
   \State Define $(i^*, q^*) = \argmax \bra{\Psi_1}\hat{\pi}_i(q)\ket{\Psi_1}$ (see Eq.~\eqref{q_projector}). 
   \State Initialize $\ket{\Psi'}=\ket{\Psi_1}$ and project qubit $i^*$ to $\ket{+_{i^*}}$.
   \If{$i^*> r$ ($i^*$ was non-stabilizer)}
    \State Optimize variationally the non-stabilizer part of $\ket{\Psi'}$.
   \EndIf
   \State Define $\hat{U}'=\hat{U}_1\hat{V}$ (see Eq.\eqref{V_unitary}).
   \end{algorithmic}
\textbf{Output:} the renormalized initial state $\ket{\Psi'}$ and the equivalent unitary $\hat{U}'$ applied to it.
\end{algorithm}

The next step to be taken depends on how $\hat{P}$ acts on $\ket{\Psi_0}$, i.e., whether it spreads over many qubits and which ones are involved. First of all, let us check if $\hat{P}$ acts trivially on the stabilizer qubits. Suppose that for $i\leq r$ the string contains the operator $\hat{p}_i\in\{\hat{X}_i,\hat{Y}_i,\hat{Z}_i\}$. If $\hat{\varphi_i}$ of Eq.\eqref{psi0} satisfies $\hat{p}_i\ket{\varphi_i}=\pm\ket{\varphi_i}$, then we can redefine $\hat{P}$ by updating $\hat{p}_i\to\pm\hat{\mathds{1}}_i$. It might occur that in this process the string $\hat{P}$ is turned into the identity (modulo a sign): this means that the projective measurement leaves the state unaffected: we can thus drop the measurement from the circuit, and the ICCR step ends here.

In case the action of the projector is non-trivial, we first perform a local change of basis to turn $\hat{P}$ into a string of $\hat{Z}$ Pauli matrices. Let $\mathcal{S}$ be the support of $\hat{P}$, i.e., the set of sites on which it does not act as the identity. We can always find suitable single-qubit Clifford rotations $\hat{q}_i$ such that
\begin{equation}
\left(\prod_{i\in\mathcal{S}}\hat{q}_i\right)\hat{P}\left(\prod_{i\in\mathcal{S}}\hat{q}\daga_i\right) = \prod_{i\in\mathcal{S}}\hat{Z}_i.
\end{equation}
The change of basis requires us to introduce a new initial state
\begin{equation}\label{psi1}  \ket{\Psi_1}=\left(\prod_{i\in\mathcal{S}}\hat{q}_i\right)\ket{\Psi_0}=\bigotimes_{i=1}^r\ket{\Tilde{\varphi}_i}\ket{\Tilde{\Phi}_\text{non-stab}^{(N-r)}}
\end{equation}
and a new unitary $\hat{U}_1 = \hat{U}_0\left(\prod_{i\in\mathcal{S}}\hat{q}\daga_i\right)$ (see Fig.~\ref{f:algo_scheme}c). Let $s = \pm 1$ be the outcome of the measurement. The previous change of basis consists in the rewriting
\begin{equation}\label{qq_decomp}
    \hat{U}_0\frac{\hat{\mathds{1}}+s\hat{P}}{2}\ket{\Psi_0} = \hat{U}_1\frac{\hat{\mathds{1}}+s\prod_{i\in\mathcal{S}}\hat{Z}_i}{2}\ket{\Psi_1}.
\end{equation}
Notice that the outcome is set to $s=+1$ for postselected measurement originated by $T$ gadgets, whereas it must be sampled randomly according to the Born rule by measuring the expectation value $\bra{\Psi_1}\prod_{i\in\mathcal{S}}\hat{Z}_i\ket{\Psi_1}$.

The fact that the projective measurement fixes the parity $\prod_{i\in\mathcal{S}}\hat{Z}_i$ to $s$ allows us to represent its action on $\ket{\Psi_1}$ in terms of Clifford unitaries acting on a renormalized state. If we expand $\ket{\Psi_1}$ in the computational basis as
\begin{equation}
    \ket{\Psi_1} =\sum_{\sigma_1,\dots,\sigma_N=\pm 1} a_{\sigma_1 \dots \sigma_N} \ket{\sigma_1,\dots,\sigma_N} \, ,
\end{equation}
the projector destroys all terms in the linear combination that do not fulfill the constraint $\prod_{i\in\mathcal{S}}\sigma_i = s$. The constraint can be thought as reducing by one the free parameters in $\mathcal{S}$. We can mimick the effect of the projective measurement $\prod_{i\in\mathcal{S}}\hat{Z}_i$ with unitary operators by picking a qubit $i^*\in \mathcal{S}$ to be the target qubit of a CNOT gates cascade designed to enforce the parity constraint, as shown in Fig.~\ref{f:algo_scheme}d. While obviously any qubit in $\mathcal{S}$ can be picked as the target, the optimal choice consists of the qubit of $\ket{\Psi_1}$ that is closest to an eigenstate of $\hat{X}$ or $\hat{Y}$~\footnote{As we shall see later, replacing the measurement with unitary gates can require a redefinition of the initial state of the circuit, and possibly an increase in its complexity. If the target qubit is in an eigenstate of $\hat{X}$ or $\hat{Y}$ the redefinition is trivial, and it does not bear any growth of complexity.}. 
Formally, we require that
\begin{equation}
    (i^*, q^*) = \argmax_{i\in\mathcal{S},\,q\in\{0,1,2,3\}} 
    \bra{\Psi_1}\hat{\pi}_i(q) \ket{\Psi_1} \, ,
\end{equation}
where
\begin{equation}\label{q_projector}
    \hat{\pi}_i(q) = \ket{\psi_i(q)}\bra{\psi_i(q)}
\end{equation}
is a projector onto the local state $\ket{\psi_i(q)} = (\ket{+_i}+i^q\ket{-_i})/\sqrt{2}$ ($q=0,1,2,3$) that parameterizes $\hat{X}, \hat{Y}$ eigenstates on site $i$. Qubit $i^*$ is the optimal target, whereas $q^*$ determines which eigenstate of $\hat{X}_{i^*}$, $\hat{Y}_{i^*}$ it is closest to. If $\mathcal{S}$ overlaps with the stabilizer qubits we would find $i^*\leq r$~\footnote{We previously removed from $\mathcal{S}$ all sites $i\leq r$ corresponding to eigenstates of $\hat{Z}$. Hence, any $i\in\mathcal{S}$ with $i\leq r$ must correspond to an eigenstate of $\hat{X}$ or $\hat{Y}$.}. In this case the identity
\begin{equation}\label{final_replacement}
    \frac{\hat{\mathds{1}}+s\prod_{i\in\mathcal{S}}\hat{Z}_i}{2}\ket{\Psi_1} = \mathcal{N}\hat{V}\ket{\Psi'} \, ,
\end{equation}
holds. Here $\mathcal{N}$ is the norm of the left-hand side, $\hat{V}$ is the Clifford operator
\begin{equation}\label{V_unitary}
    \hat{V} = \hat{S}^{q^*}_{i^*} \left(\prod_{i\in\mathcal{S}\setminus\{i^*\}}\hat{\mathrm{CX}}_{i\to i^*}\right)\hat{X}_{i^*}^{\frac{1-s}{2}},
\end{equation}
and the state
\begin{equation}\label{psi_prime_exact}
\ket{\Psi'}=\bigotimes_{i=1}^{i^*-1}\ket{\Tilde{\varphi}_i}\ket{+_{i^*}}\bigotimes_{i=i^*+1}^r\ket{\Tilde{\varphi}_i}\ket{\Tilde{\Phi}_\text{non-stab}^{(N-r)}}
\end{equation}
is equal to $\ket{\Psi_1}$ apart from having qubit $i^*$ set to $\ket{+}$.

We successfully replaced the initial circuit with a measurementless one, characterized by the renormalized initial state $\ket{\Psi'}$ and the new Clifford unitary $\hat{U}'=\hat{U_1}\hat{V}$. Notice that the non-stabilizer part of $\ket{\Psi'}$ is connected to that of the starting state $\ket{\Psi_0}$ by single-qubit rotations, and we conclude that the projective measurement does not modify the magic of the state.

A different scenario unfolds when $i^*>r$, in which case the projective measurement acts only on the non-stabilizer part of the state and impacts it in a non-trivial way. A final form like in Eq.~\eqref{final_replacement} can still be obtained, but $\ket{\Psi'}$ will grow in complexity as compared to $\ket{\Psi_1}$ (see App.~\ref{a:proj_replacement}). Even in the special case in which $\ket{\Phi_\text{non-stab}^{(N-r)}}$ of Eq.~\eqref{psi0} is a tensor product of single-particle states, the action of the measurement typically results in an entangled renormalized initial state $\ket{\Psi'}$. The only approximation of this algorithm is made here: to keep the problem tractable, we use a variational approximation of $\ket{\Psi'}$ and we optimize it within a given class of states to minimize the distance between left and right-hand sides of Eq.~\eqref{final_replacement}. In Sec.~\ref{s:numerics} we implement an MPS approximation, showing that it performs well even with small bond dimension $\chi$ (as compared to those needed to explicitly represent the full state). Details on the optimization can be found in App.~\ref{a:variational}. Proceeding in this way, we finally have
\begin{equation}\label{psi_final_magic}
    \ket{\Psi'}\approx \bigotimes_{i=1}^r \ket{\Tilde{\varphi}_i}\ket{+_{i^*}} \ket{\Tilde{\Phi}_\text{non-stab}^{(N-r-1)}},
\end{equation}
where the state of non-stabilizer qubits $\ket{\Tilde{\Phi}_\text{non-stab}^{(N-r-1)}}$ is obtained variationally. Notice that Eq.~\eqref{psi_final_magic} is completely analogous to Eq.~\eqref{psi_prime_exact}: the stabilizer states are the same as $\ket{\Psi_1}$, and the target qubit is set to $\ket{+_{i^*}}$. Notice however that in this case the stabilizer rank has been explicitly increased by $1$, manifestly changing the magic of the state.

\subsection{Computational cost}\label{ss:cost}
Here we discuss the efficiency of each step of the ICCR technique. The overall cost depends on the choice of the variational ansatz: in the following, we assume the simplest product-state approximation. In our implementations, we store in memory the stabilizer tableaus of unitary gates. Assuming that at the beginning of the procedure $\hat{U}_0$ consists of a single brick layer of Fig.~\ref{f:circuit}, the cost to compose $N/2$ two-qubit gates is $\mathcal{O}(N^2)$. The evaluation of Eq.~\eqref{P_string} is performed in $\mathcal{O}(N)$ operations using stabilizer-state methods~\cite{Gidney_2021}. Its simplification by checking how it acts on the states $\ket{\varphi_i}$ is also achieved in at most $\mathcal{O}(N)$ operations. Next, Eq.~\eqref{psi1} involves applying $|\mathcal{S}|$ single-qubit unitaries, for a cost that scales as $\mathcal{O}(|\mathcal{S}|)\leq\mathcal{O}(N)$. Evaluating $\bra{\Psi_1}\prod_{i\in\mathcal{S}}\hat{Z}_i\ket{\Psi_1}$ and finding $i^*$ and $q^*$ also have the same cost. The evaluation of $\hat{U}'$ amounts to composing $\hat{U}$ with $\mathcal{O}(|\mathcal{S}|)$ one or two-qubit gates, which costs $\mathcal{O}(N|\mathcal{S}|)\leq\mathcal{O}(N^2)$ operations. Finally, the cost of the optimization required to determine Eq.~\eqref{psi_final_magic} is $\mathcal{O}((N-r)^2\chi^3)\leq \mathcal{O}(N^2\chi^3)$, as shown in App.~\ref{a:variational}. In summary, the overall maximal cost of the algorithm is $\mathcal{O}(N^2\chi^3)$ per layer and per projective measurement. Assuming a total of $M$ measurements, this brings the overall complexity to $\mathcal{O}(N^2 \chi^3 M)$.

For circuits involving a number $n_T$ of $T$ gates a basic way to treat the problem is to replace each of them with a $T$ gadget by introducing $n_T$ ancilla qubits. The total number of qubits to simulate grows, thus increasing the complexity to $\mathcal{O}((N+n_T)^2)$ per layer and per projective measurement in the worst case scenario. While the simulation is still efficient, it can become quite expensive for Clifford circuits doped with many $T$ gates. In App.~\ref{a:T_gates} we show that there is a better way of proceeding that keeps the computational cost at $\mathcal{O}(N^2)$. The idea of this alternative approach is that the final state of the ancilla qubit of the $T$ gadged is disentangled from the physical qubits, making it possible to remove its presence from both the effective initial state and the stabilizer tableau of the stored unitary gate. As a consequence, both $\ket{\Psi}$ and $\hat{U}$ are restored to being defined on $N$ qubits.

\subsection{Discussion on the approximation and its generalization}\label{ss:approx}
The ICCR algorithm provides a renormalization flow of the initial state $\ket{\Psi}$, which is however expensive to keep track of exactly and is thus approximated by restricting it to a variational class of states. A rather drastic approach is to take the latter as the class of product states, which has the advantage of simplifying the calculation of the SREs of Eq.~\eqref{sres} into a sum of single-qubit SREs due to the additivity property. While this approximation may seem dramatic, our numerical analysis presented in Sec.~\ref{s:numerics} shows that it already captures the qualitative properties of the dynamics of magic. The quality of the approximation is controlled by how close the target qubit $i^*$ is to the eigenstates of $\hat{X}$ and $\hat{Y}$. If the size of the support $\mathcal{S}$ in case (4.b) is large the choice of the target is broad and it is more likely to obtain a good approximation. In general, we expect errors to build up as the procedure is iterated, reasonably affecting the performance of the ICCR algorithm at long times.

Beyond this, one can play with the variational class replacing product states with MPSs with fixed bond dimension $\chi$. Though naively analogous, this is very different from performing a full tensor network simulation of the problem, as our goal is still to avoid computing the dynamics. The idea is that while the initial state flow may build up some entanglement, it does not produce as much as the circuit evolution, and thus it can be efficiently captured with a low-$\chi$ MPS. The calculation of the SREs becomes more involved, but can still be done efficiently for low enough $\chi$ using the perfect sampling technique of Ref.~\cite{Lami_2023}, which involves extracting samples of Pauli strings and leveraging the outcomes to form statistical estimators. Throughout this manuscript, we sample $10^3$ Pauli strings per SRE measurement. This generalization of the approximation also allows to verify the quality of the estimation by checking the convergence of the SREs as the bond dimension is increased. Indeed, in Sec.~\ref{s:numerics} we show that the accuracy improves with the bond dimension used, and the SREs converge to the exact values as $\sim \chi^{-1}$.

The ICCR step can be effectively viewed as an approach for approximating a generic quantum state using an MPS evolved with a Clifford unitary. This paradigm has been recently established under the name of Clifford enhanced (or augmented) MPSs by Ref.~\cite{Lami_2024_1} in the context of quantum state designs~\cite{Ippoliti_2022,Gross_2007,mele_2024}, and by Refs.~\cite{qian2024,mello2024,huang2024,fux2024} from the perspective of increasing the numerical performance of MPS algorithms. These works suggest a way to further improve the ICCR ansatz by optimizing variationally not only the MPS in Eq.~\eqref{psi_final_magic}, but also the Clifford operator $\hat{V}$ appearing in Eq.~\eqref{final_replacement}, rather than assuming the given form of Eq.~\eqref{V_unitary}. The implementation of this generalization is left for future studies.

We expect that the ICCR estimates the stabilizer nullity very accurately, as the latter is insensitive to the specific definition of the non-stabilizer part of the state. Indeed, whether or not the rank increases by $1$ depends only on the expression of the Pauli string $\hat{P}$ of Eq.~\eqref{P_string} and on the stabilizer qubit states, and not on the details of $\ket{\Phi_\text{non-stab}^{(N-r)}}$. The nullity is also independent of the outcomes of the measurements, which are sampled from an approximation of the initial state~\footnote{The difference between the two possible measurement outcomes only amounts to the presence or absence of an $\hat{X}$ gate in Eq.~\eqref{V_unitary}. This gate, being a $\pi$ rotation, can at most affect the outcomes $s$ of the following iterations.} There is however a very specific case in which the nullity is not estimated properly. While in our framework the rank can only increase by a unit as a result of a measurement, there exist non-stabilizer states for which it grows by more~\cite{Beverland_2020}.
These processes are not captured by our framework, making our estimate of $\mathcal{\nu}$ only an upper bound. Nevertheless, when comparing our method to exact diagonalization we found these instances to be extremely rare, making us confident that the ICCR estimates the stabilizer nullity very accurately.

\section{Numerical results}\label{s:numerics}
In this section we present an example of application of the ICCR algorithm, as well as a benchmark of its performance. We consider a model of measurement-induced phase transitions considered in Refs.~\cite{Skinner_2019} and~\cite{Gullans_2020}, which consists of a 1D random Clifford circuit with projective measurements, as schematized in Fig.~\ref{f:circuit}. Differently from these works, here we are interested in the dynamics of magic when the system is initially prepared in a non-stabilizer state $\ket{\Psi_0}$. We consider
\begin{equation}\label{initial_state}
    \ket{\Psi_0} = \left[\cos(\pi/7)\ket{+}+\sin(\pi/7)\ket{-}\right]^{\otimes N},
\end{equation}
which has extensive non-stabilizerness. Neglecting the measurements, the unitary part of the dynamics forms a brick-wall circuit of depth $T$, where each two-qubit gate is picked randomly from the two-qubit Clifford group. We assume a ring geometry, so that the first and last qubits are connected by gates. Projective measurements in the $Z$ basis are performed between layers of unitary operators with a density $0<p<1$, namely, each spin has a probability $p$ of being measured after being subject to a unitary gate. The measurement outcomes are picked randomly according to the Born rule. As we show below, we observe that the magic features a measurement-induced transition when varying $p$.

\subsection{Magic phase transition}
\begin{figure}[t!]
    \centering
    \includegraphics[width=\columnwidth]{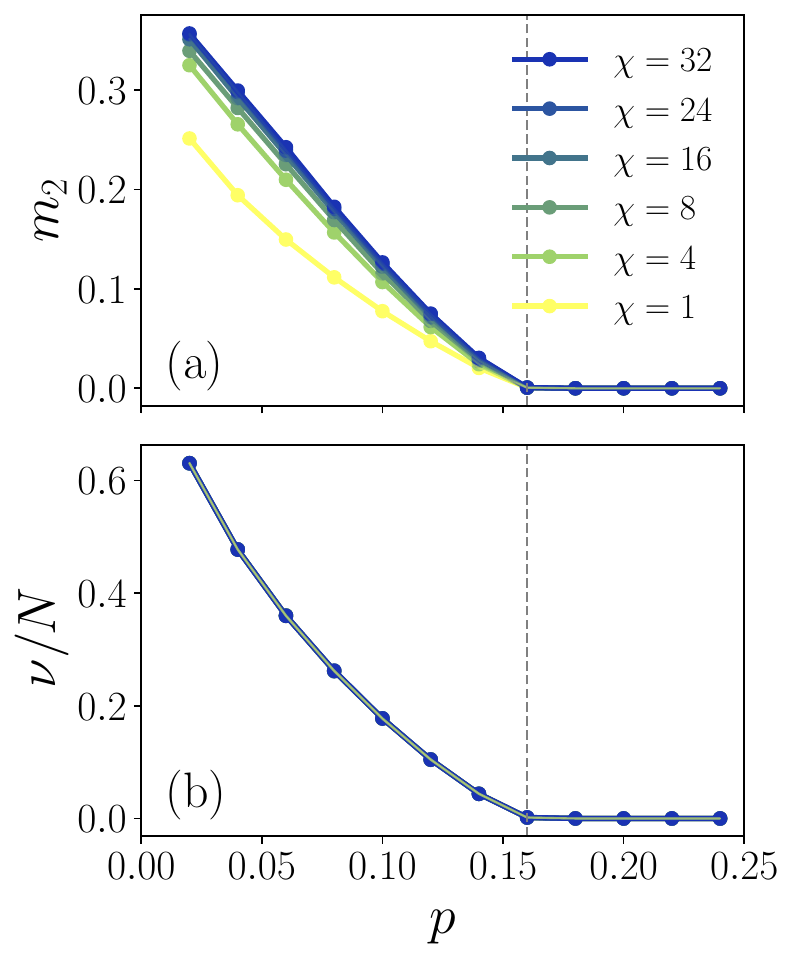}
    \caption{Long-time magic densities (a) $m_2$ and (b) $\nu/N$, as functions of the measurement rate $p$, computed with the ICCR algorithm using different bond dimensions $\chi$. Data for $N=1000$, $T=2000$, averaged over $200$ realizations. All magic densities vanish for $p>p_c\approx 0.16$, whereas the system still features extensive non-stabilizerness below $p_c$.}
    \label{f:phase_trans}
\end{figure}

We employ the ICCR method to study the dynamics of non-stabilizerness in the monitored Clifford circuit described above. This model is known to feature a measurement-induced phase transition~\cite{Skinner_2019,Gullans_2020} in both entanglement and mixed-state purification dynamics with critical point at $p_c\approx 0.16$. To track the renormalization of the initial state, we use MPSs with bond dimension $\chi$ as the variational class. Figure~\ref{f:phase_trans} shows our numerical results for the densities of second SRE $m_2=M_2/N$ and nullity $\nu/N$ for a system of $N=1000$ qubits and depth $T=2N$. We distinguish two phases separated by a critical measurement density $p_c$ that coincides with that observed in previous works. For $p>p_c$, the measurements purify the magic of the system, leaving it in an almost-stabilizer state. In contrast, an extensive non-stabilizerness survives for $p<p_c$, meaning that the state is still non-trivial at times of order $\mathcal{O}(N)$.
\begin{figure}[t!]
    \centering
    \includegraphics[width=\columnwidth]{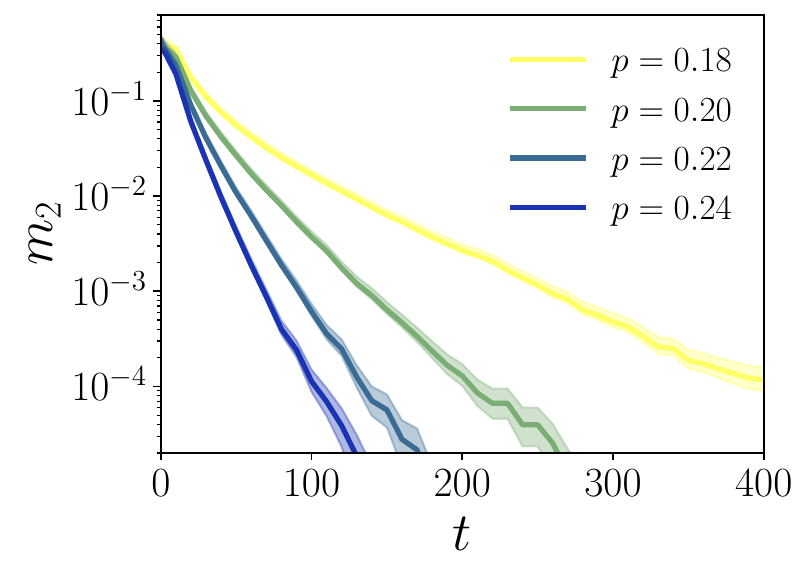}
    \caption{Dynamics of the second SRE density $m_2$ for $p>p_c$. Data for $N=1000$ averaged over $200$ realizations. We use $\chi=32$, although other choices of $\chi$ produce qualitatively analogous behavior. The magic relaxes to zero in a finite time with an exponential time dependence.}
    \label{f:dyn_fast}
\end{figure}

As the bond dimension $\chi$ is increased, the renormalization of the initial state is followed more accurately, and thus we expect the non-stabilizerness estimates to converge to the exact results for $\chi\gg 1$. Interestingly, we notice that $m_2$ grows monotonically as $\chi$ is highered, indicating that computing the SRE at finite bond dimension systematically underestimates the true value. This numerical observation suggests that the ICCR method provides lower bounds to the magic measures, even though we were unable to prove it rigorously. In contrast, as argued previously, the stabilizer rank is independent of $\chi$.

\begin{figure}[t!]
    \centering
    \includegraphics[width=\columnwidth]{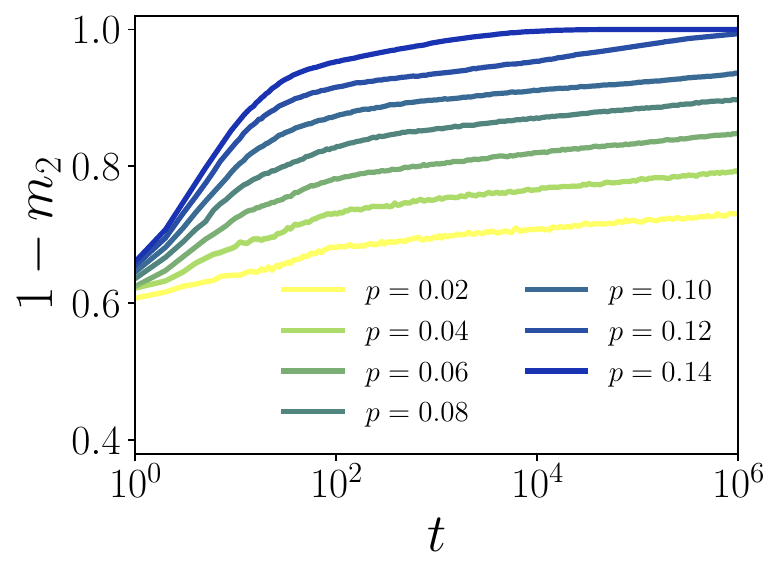}
    \caption{Dynamics of $1-m_2$, the complementary of the second SRE density, for $p<p_c$. Data for $N=140$ averaged over $100$ realizations. We use $\chi=16$, although other choices of $\chi$ produce qualitatively analogous behavior. Asymptotically, $1-m_2$ grows logarithmically in time.}
    \label{f:dyn_slow}
\end{figure}

The two phases are better distinguished by looking at how magic evolves as a function of the discrete time $t$ of the circuit. For $p>p_c$, we show in Fig.~\ref{f:dyn_fast} that the steady state with vanishing non-stabilizerness is reached exponentially fast, with a typical relaxation time that depends on $p$ and we find to be independent of $N$. In contrast, Fig.~\ref{f:dyn_slow} shows the case of $p<p_c$, which features very slow dynamics given by $M_2(t)\approx N-A \ln t$ at long times. We observe that the prefactor $A$ of the logarithmic growth is independent of $N$, and thus it takes an exponentially long time for $m_1$ to relax to zero.

These results are consistent with the numerical investigation presented in Ref.~\cite{Gullans_2020}, which considers the evolution of the von Neumann entropy $S(\hat{\rho}(t))$ of a density matrix $\hat{\rho}(t)$ initially prepared in $\hat{\rho}(0) = \hat{\mathds{1}}/2^N$. For $p>p_c$ they observe that the entropy density relaxes to zero in a finite, $N$-independent time, whereas it remains finite for exponentially long times at small $p<p_c$, precisely as we find for magic measures. Indeed, the authors show that the entropy $S(\hat{\rho})$ actually coincides with the stabilizer nullity $\nu$ of $\hat{\rho}$ for the problem they consider. Our investigation thus confirms the findings of Ref.~\cite{Gullans_2020}, and generalizes it to the case of a pure non-stabilizer initial state.

\subsection{Benchmark of the accuracy}
\begin{figure}[t!]
    \centering
    \includegraphics[width=\columnwidth]{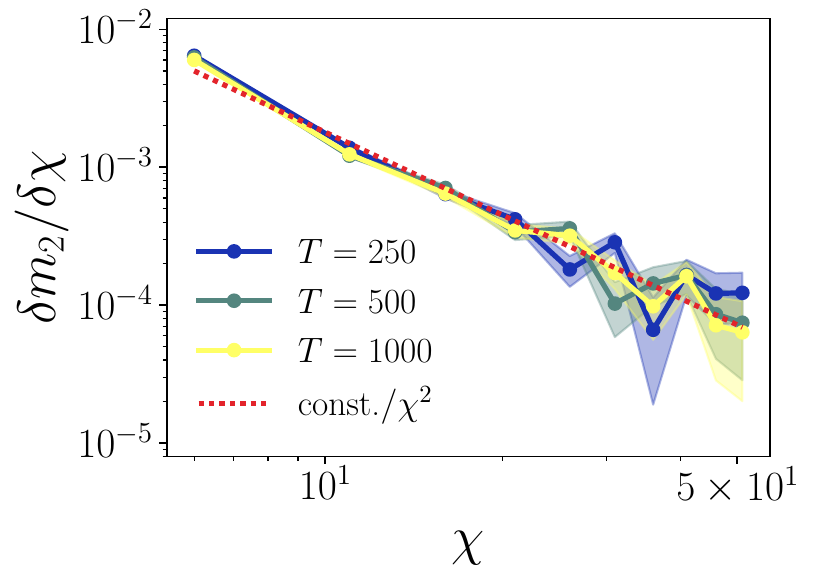}
    \caption{Estimate of $\delta m_2/\delta \chi$ evaluated at different cirucit depths $T$. Data for $N=140$ and $p=0.1$ averaged over $2500$ realizations.}
    \label{f:convergence}
\end{figure}

We now investigate the rate of convergence of the SREs estimates as the bond dimension $\chi$ is increased. This allows us to quantify the bond dimension (and thus the amount of computational resources) needed to achieve a given accuracy. We focus on the phase $p<p_c$, because as shown above in the other region the SREs converge quickly with $\chi$. We consider a system size of $N=500$, a measurement density $p=0.1$, and various possible depths $T=250,500,1000$. For each of these, we evaluate $m_2(\chi)$ at different bond dimensions $\chi$. We then estimate the variation $\delta m_2/\delta \chi \approx [m_2(\chi)-m_2(\chi-5)]/5$, using a step $\delta \chi = 5$ is to mitigate numerical error. Figure~\ref{f:convergence} shows our numerical results. We observe a power-law decay $\delta m_2/\delta \chi \sim \chi^{-2}$. This implies that the cumulative error $m_2(\infty)-m_2(\chi)$ is expected to scale as $\chi^{-1}$, meaning that the bond dimension required to reach a given accuracy $\epsilon$ grows as $\epsilon^{-1}$. We also point out that $\delta m_2/\delta \chi<0$, which further indicates that finite-$\chi$ estimates $m_2(\chi)$ are lower bounds of the exact value.

\begin{figure}[t!]
    \centering
    \includegraphics[width=\columnwidth]{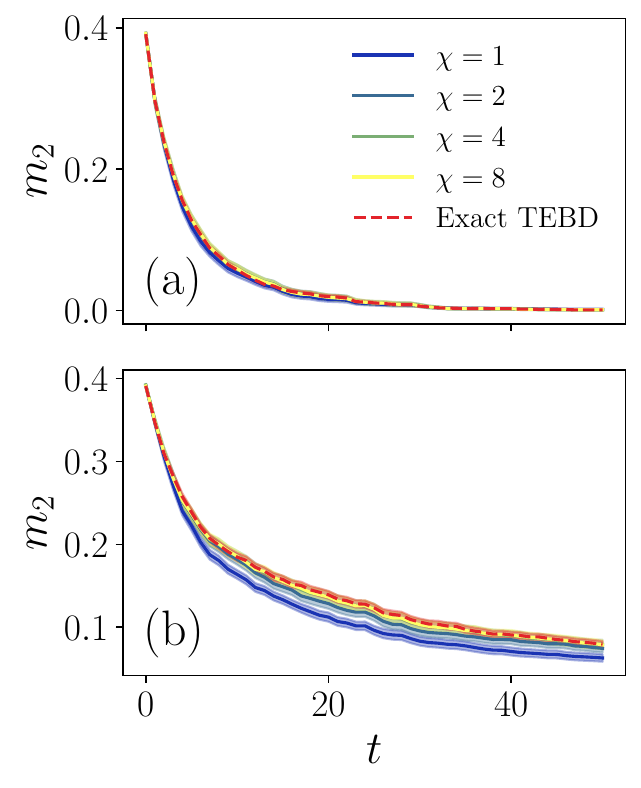}
    \caption{Average SRE density $m_2$ computed using the ICCR algorithm with various bond dimensions $\chi$ (solid lines) and exact TEBD (dashed red line). Data for $N=20$ and (a) $p=0.22$, (b) $p=0.1$, averaged over $50$ random realizations.}
    \label{f:MPS_vs_ICCR}
\end{figure}
To verify the correctness of the ICCR approach, we now compare the values of the stabilizer Renyi entropies obtained using this technique with those obtained through standard tensor network methods. To this purpose, we perform an MPS simulation without resorting to the state renormalization techniques explained above, simply by contracting the quantum circuit using the time-evolving block decimation (TEBD) algorithm. We consider a system of size $N=20$ and study the evolution of the SRE in a circuit of depth $T=50$ with open boundary conditions. Importantly, at this system size we are able to follow the state exactly without truncating the TEBD bond dimension. Projective measurements were implemented in the MPS framework using standard techniques~\cite{Stoudenmire_2010,Ferris_2012}. The SREs of this simulation are evaluated using the perfect sampling algorithm~\cite{Lami_2023}, using a sampling of $10^3$ Pauli strings.

In Fig.~\ref{f:MPS_vs_ICCR}, we present the values of the SRE density $m_2$ obtained with the two approaches~\footnote{For this comparison, we adopted the same trajectories (i.e., the same choices of Clifford gates and measurement outcomes) in the two methods.}, for two different values of the projective measurements density $p=0.1$ and $p=0.22$. We observe an excellent agreement in the phase $p>p_c$ already with a product-state ansatz ($\chi=1$), while in the phase $p<p_c$ we require a larger bond dimension to obtain matching values. This shows that the ICCR estimates are unbiased, and they will converge to the exact SREs when using a large enough bond dimension. As observed previously, also in this case the values obtained at small $\chi$ appear to lower bound the exact ones.

\begin{figure}[t!]
    \centering
    \includegraphics[width=0.8\columnwidth]{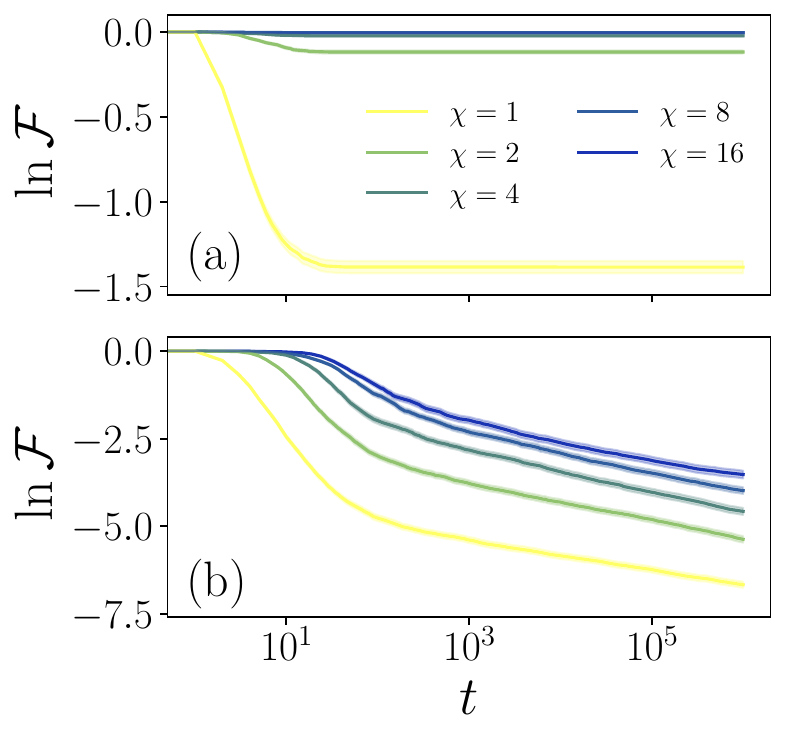}
    \caption{Dynamics of the logarithmic cumulative fidelity $\ln \mathcal{F}$ in the two phases: (a) $p=0.22$, and (b) $p=0.1$. Data for $N=140$ averaged over $200$ realizations.}
    \label{f:fidelity}
\end{figure}

The different accuracy of the ICCR method in the two phases is reflected qualitatively in the error made by the approximation of Eq.~\eqref{psi_final_magic}. A possible way to quantify it is by keeping track of the overlap between exact and approximate states at each ICCR iteration. Let $\ket{\Psi'_{n,\text{exact}}}$ be the state that solves Eq.~\eqref{final_replacement} at the $n$th iteration of the ICCR algorithm, and let $\ket{\Psi'_{n,\text{approx}}}$ be its approximation obtained from Eq.~\eqref{psi_final_magic} (see App.~\ref{a:variational}). We define the squared overlap 
\begin{equation}\label{fidelity}
    f_n = |\braket{\Psi'_{n,\text{exact}}}{\Psi'_{n,\text{approx}}}|^2.
\end{equation}
Denoting by $M(t)$ be the number of ICCR iterations performed up to a given circuit depth $t$, we introduce the cumulative fidelity
\begin{equation}
    \mathcal{F}(t) = \prod_{n=1}^{M(t)} f_n
\end{equation}
to estimate the drift of the iteratively approximated state from the exact one. In Fig.~\ref{f:fidelity} we show its dynamics for measurement rates $p$ in the two phases. As expected, a larger bond dimension $\chi$ implies more variational freedom in the approximation of the state, and as a result the fidelity is higher. The time dependence of $\ln \mathcal{F}$ appears to be logarithmic in time, indicating a power-law decay of the fidelity. For $p=0.22>p_c$ and $\chi=1$, it saturates quickly to a constant $\mathcal{F}\approx 0.25$, which is very large especially considering the system size of $N=140$; as $\chi$ is increased, the fidelity rapidly approaches $1$. In contrast, for $p=0.1<p_c$ we observe two algebraic regimes: at short times the decay is faster, whereas it features a smaller exponent as longer times. The fidelity decreases much more in this phase, consistently with the worse performance of the ICCR method found in Fig.~\eqref{f:MPS_vs_ICCR}. Based on this, we speculate that the fidelity can be used to quantify the quality of the approximation of the ICCR method.

\section{Conclusions}\label{s:conclusions}
In this work we propose the new ICCR algorithm to face the challenging problem of evaluating magic in quantum circuits. The main advantage of this technique is to avoid entirely to compute the dynamics of the state explicitly, resulting in a computational cost that scales only polynomially in the number of degrees of freedom. This result is achieved by an iterative manipulation of the structure of the circuit and of the initial state, exploiting the property that Clifford unitary gates preserve the non-stabilizerness. We showcase an application of ICCR by studying a 1D circuit of up to $N=1000$ involving projective measurement, showing that it exhibits a measurement-induced phase transition in the rate of magic purification. The estimated error of our technique scales as the inverse of the bond dimension $\chi$ of the MPS ansatz used to capture the initial state, and we observe numerically that results at finite $\chi$ lower bound the exact magic values.

We believe that the ICCR is a powerful and versatile tool to investigate a variety of problems involving non-stabilizerness dynamics, such as those considered in Refs.~\cite{Fux_2023,Bejan_2023,Niroula_2023,Turkeshi_2023} on hybrid quantum circuits, at very large system sizes. We expect the ICCR method to reproduce the results of these works with lower computational cost, allowing for an in-depth study of the phase transitions observed. An interesting question left for future studies would be to investigate the performance of ICCR combined with other classes of states for the variational approximation. Finally, we remark that the ICCR method is independent of the geometry of the gates, and thus it could be used to unveil the behavior of magic in higher dimensional or long-range circuits whose simulation with standard simulation techniques is highly challenging.

\section{Acknowledgements}
\label{sec:acknowledgements}
A.P. thanks E. Tirrito, A. Hamma, D. Iannotti, and G. Esposito for useful discussions. This work has been supported by the PNRR MUR project PE0000023-NQSTI, the Quantera project ``SuperLink'', and the PRIN 2022 (2022R35ZBF) - PE2 - ``ManyQLowD''. G.L. was partially supported by ANR-22-CPJ1-0021-01.

\appendix

\section{Replacing a projector with unitary gates}\label{a:proj_replacement}
We prove the identities of Eqs.~\eqref{final_replacement},~\eqref{V_unitary},~\eqref{psi_prime_exact}, and we provide the formal exact expression of $\ket{\Psi'}$ of Eq.~\eqref{psi_final_magic}.

First, assume we can find $i^*\leq r$ with $\ket{\Tilde{\varphi}_{i^*}}=(\ket{+}+i^{q^*}\ket{-})/\sqrt{2}$ (cf. Eq.~\eqref{psi1}). Let us perform a computational-base expansion of $\ket{\Psi_1}$. We have
\begin{equation}
\begin{split}
    \ket{\Psi_1} &= \sum_{\sigma_1,\dots,\sigma_N=\pm 1} a_{\sigma_1,\dots,\sigma_N}\ket{\sigma_1,\dots,\sigma_N}\\
    &= \sum_{\sigma_i\;:\;i\in\mathcal{S}} \ket{\sigma_i\;:\;i\in\mathcal{S}} \sum_{\sigma_i\;:\; i\notin\mathcal{S}}a_{\sigma_1,\dots,\sigma_N}\ket{\sigma_i\;:\;i\notin\mathcal{S}}\\
    &= \sum_{\sigma_i\;:\;i\in\mathcal{S}} \ket{\sigma_i\;:\;i\in\mathcal{S}} A_{\{\sigma_i\;:\;i\in\mathcal{S}\}}\ket{\psi_{\{\sigma_i\;:\;i\in\mathcal{S}\}}},
\end{split}
\end{equation}
where we isolated the contribution of spins contained in $\mathcal{S}$ by grouping all others in the states $\ket{\psi_{\{\sigma_i\;:\;i\in\mathcal{S}\}}}$. The latter are normalized to $1$, and thus the amplitudes $A_{\{\sigma_i\;:\;i\in\mathcal{S}\}}$ appear in the expansion. Let us introduce a string $\ssigma = \{\sigma_i\;:\;i\in\mathcal{S}\setminus\{i^*\}\}$ of length $|\mathcal{S}|-1$ containing all qubits in $\mathcal{S}$ except for the one at $i^*$. $\ket{\Psi_1}$ can be rewritten compactly as
\begin{equation}\label{psi1_first_expansion}
    \ket{\Psi_1} = \sum_{\ssigma}\sum_{\sigma_{i^*}} A_{\ssigma,\sigma_{i^*}} \ket{\ssigma}\ket{\sigma_{i^*}}\ket{\psi_{\ssigma,\sigma_{i^*}}}.
\end{equation}
Using the fact that qubit $i^*$ is in the known product state $\ket{\Tilde{\varphi}_{i^*}}$, the amplitudes factorize as $A_{\ssigma,\sigma_{i^*}}=\Tilde{A}_{\ssigma}(i^{q^*})^\frac{1-\sigma_{i^*}}{2}/\sqrt{2}$ and the states $\ket{\psi_{\ssigma,\sigma_{i^*}}}$ are independent of $\sigma_{i^*}$, thus yielding
\begin{equation}\label{psi1_expansion}
    \ket{\Psi_1} = \sum_{\ssigma} \Tilde{A}_{\ssigma}  \ket{\ssigma}\sum_{\sigma_{i^*}}\frac{(i^{q^*})^\frac{1-\sigma_{i^*}}{2}}{\sqrt{2}}\ket{\sigma_{i^*}}\ket{ \psi_{\ssigma}}.
\end{equation}
We now evaluate the action of the projector $(\hat{\mathds{1}}+s\prod_{i\in\mathcal{S}}\hat{Z}_i)/2$ on this state. The projector preserves all components of the superposition satisfying $\prod_{i\in\mathcal{S}}\sigma_i = s$, and destroys all others. As a consequence, defining $\Bar{\sigma} = \prod_{i\in\mathcal{S}\setminus \{i^*\}} \sigma_i$, for any given $\ssigma$ only states with $\sigma_{i^*} = s\Bar{\sigma}$ are not annihilated, yielding
\begin{equation}
    \frac{\hat{\mathds{1}}+s\prod_{i\in\mathcal{S}}\hat{Z}_i}{2}\ket{\Psi_1} = \sum_{\ssigma} \Tilde{A}_{\ssigma} \frac{(i^{q^*})^\frac{1-s\Bar{\sigma}}{2}}{\sqrt{2}}\ket{\ssigma}\ket{s\Bar{\sigma}}\ket{ \psi_{\ssigma}}.
\end{equation}
The state $\ket{s\Bar{\sigma}}$ can be achieved by preparing qubit $i^*$ in $\ket{s}$ and controlling it with $|\mathcal{S}|-1$ CNOT gates applied from the spins in $\mathcal{S}\setminus \{i^*\}$. The phase $(i^{q^*})^\frac{1-s\Bar{\sigma}}{2}$ is reproduced with the gate $\hat{S}^{q^*}_{i^*}$, namely,
\begin{equation}
\begin{split}
    \frac{\hat{\mathds{1}}+s\prod_{i\in\mathcal{S}}\hat{Z}_i}{2}\ket{\Psi_1} =& \frac{1}{\sqrt{2}}\hat{S}^{q^*}_{i^*}\left(\prod_{i\in\mathcal{S}\setminus \{i^*\}}\hat{\mathrm{CX}}_{i\to i^*}\right) \\
    &\cdot \sum_{\ssigma} \Tilde{A}_{\ssigma}\ket{\ssigma}\ket{s}\ket{ \psi_{\ssigma}}.
\end{split}
\end{equation}
Notice that the state on the last line of the previous equation is closely related to Eq.~\eqref{psi1_expansion}, as it differs only by having $\ket{s}$ instead of $\ket{\varphi_{i^*}}$ on site $i^*$. Using the identity $\ket{s} = \hat{X}^\frac{1-s}{2} \ket{+}$, we finally obtain Eq.~\eqref{final_replacement} with $\mathcal{N} = 1/\sqrt{2}$.

Let us now assume that the target qubit $i^*$ is non-stabilizer. The simplification of Eq.~\eqref{psi1_first_expansion} to Eq.~\eqref{psi1_expansion} is no longer possible, the amplitudes $A_{\ssigma,\sigma_{i^*}}$ do not factorize, and the states $\ket{\psi_{\ssigma,\sigma_{i^*}}}$ depend also on $\sigma_{i^*}$. Acting with the projector we obtain
\begin{equation}
\begin{split}
    \frac{\hat{\mathds{1}}+s\prod_{i\in\mathcal{S}}\hat{Z}_i}{2}\ket{\Psi_1} &= \sum_{\ssigma} A_{\ssigma,s\Bar{\sigma}} \ket{\ssigma}\ket{s\Bar{\sigma}}\ket{ \psi_{\ssigma,s\Bar{\sigma}}}.
\end{split}
\end{equation}
Proceeding as in the previous case, we achieve the final form of Eq.~\eqref{final_replacement}, but the state $\ket{\Psi'}$ takes the non-trivial form
\begin{equation}\label{psi_final_exact}
    \ket{\Psi'} = \mathcal{N}\sum_{\ssigma} A_{\ssigma,s\Bar{\sigma}}(i^{-q^*})^{s\Bar{\sigma}} \ket{\ssigma}\ket{+}\ket{\psi_{\ssigma,s\Bar{\sigma}}},
\end{equation}
where $\mathcal{N}$ is a normalization prefactor (not necessarily equal to $\sqrt{2}$ as in the previous case). We see explicitly that the rank increases by one, as the initially non-stabilizer target qubit $i^*$ is brought to $\ket{+}$. Instead, the state of the other qubits increases in complexity, because in general it develops entanglement even when starting from a product state configuration.

\section{Variational approximation of $\ket{\Psi'}$}\label{a:variational}
As mentioned in the main text, the ICCR procedure can lead to a growth of complexity of the initial state, which requires to approximate the non-stabilizer part of Eq.~\eqref{psi_final_magic}. When using an MPS ansatz for $\ket{\Psi'}$, the approximation can be performed variationally with a standard MPS compression algorithm. In the following, we discuss in detail how this procedure is implemented for the numerical simulations presented in Sec.~\ref{s:numerics}.

As mentioned in the main text, the only case in which the non-stabilizer part $\ket{\Tilde{\Phi}^{(N-r)}_\text{non-stab}}$ of the state $\ket{\Psi_1}$ (cf. Eq.~\eqref{psi1}) must be updated is when the support $\mathcal{S}$ is entirely contained in the range $[r+1,\dots,N]$. As a consequence, we can neglect the stabilizer qubits in Eq.~\eqref{final_replacement} and consider only the last $N-r$ ones. First, we assume that $\ket{\Tilde{\Phi}^{(N-r)}_\text{non-stab}}$ is given as an MPS
\begin{equation}
\begin{split}
    \ket{\Tilde{\Phi}^{(N-r)}_\text{non-stab}} &= \sum_{\sigma_{r+1},\dots,\sigma_N=\pm 1} \mathbb{A}_{r+1}^{\sigma_{r+1}}\dots \mathbb{A}_N^{\sigma_N} \ket{\sigma_{r+1},\dots,\sigma_N} \\
    &= 
\begin{tikzpicture}[baseline={([yshift=-0.5ex]current bounding box.center)}, scale=0.7]
\Vertex[x=0,color=orange,opacity=.7,shape=rectangle,label=$\mathbb{A}_{r+1}$,fontscale=0.75]{r+1} 
\Vertex[x=1.25,color=orange,opacity=.7,shape=rectangle,label=$\mathbb{A}_{r+2}$,fontscale=0.75]{r+2} 
    \Vertex[x=2.5,shape=rectangle,style={color=white},label=$\dots$]{dots}    \Vertex[x=3.75,color=orange,opacity=.7,shape=rectangle,label=$\mathbb{A}_N$,fontscale=0.75]{N} 
    \Edge(r+1)(r+2)
    \Edge(r+2)(dots)
    \Edge(dots)(N)
    \Vertex[x=0,y=1.25,shape=rectangle,style={color=white},label=$\sigma_{r+1}$]{sr+1}
    \Vertex[x=1.25,y=1.25,shape=rectangle,style={color=white},label=$\sigma_{r+2}$]{sr+2}
    \Vertex[x=3.75,y=1.25,shape=rectangle,style={color=white},label=$\sigma_N$]{sN}
    \Edge(r+1)(sr+1)
    \Edge(r+2)(sr+2)
    \Edge(N)(sN)
\end{tikzpicture}\;\; ,
\end{split}
\end{equation}
where we introduced standard tensor notation. In the previous equation, each $\mathbb{A}_i$ is a $\chi \times 2 \times \chi$ tensor, except for the boundary matrices $\mathbb{A}_{r+1}$ and $\mathbb{A}_N$ that are $2\times\chi$ and $\chi \times 2$ tensors, respectively. The exact state that we want to approximate is given by
\begin{equation}
    \ket{\Psi_\text{exact}} = \frac{1}{\mathcal{N}} \frac{\hat{\mathds{1}} +s\prod_{i\in \mathcal{S}}\hat{Z}_i}{2}\ket{\Tilde{\Phi}^{(N-r)}_\text{non-stab}}.
\end{equation}
The variational ansatz we use is 
\begin{equation}
    \ket{\Psi_\text{approx}} = \hat{V}\left(\ket{+_{i^*}}\otimes\ket{\Tilde{\Phi}^{(N-r-1)}_\text{non-stab}}\right),
\end{equation}
where
\begin{equation}\label{psi_out}
    \ket{\Tilde{\Phi}^{(N-r-1)}_\text{non-stab}} =
\begin{tikzpicture}[baseline={([yshift=-0.5ex]current bounding box.center)}, scale=0.7]
    \Vertex[x=0,color=blue,opacity=.4,shape=rectangle,label=$\mathbb{B}_{r+1}$,fontscale=0.75]{r+1} 
    \Vertex[x=1.25,shape=rectangle,style={color=white},label=$\dots$]{dots}
    \Vertex[x=2.5,color=blue,opacity=.4,shape=rectangle,label=$\mathbb{B}_{i^*-1}$,fontscale=0.75]{i*-1}
    \Vertex[x=3.75,color=blue,opacity=.4,shape=rectangle,label=$\mathbb{B}_{i^*+1}$,fontscale=0.75]{i*+1}
    \Vertex[x=5,shape=rectangle,style={color=white},label=$\dots$]{dots2}
    \Vertex[x=6.25,color=blue,opacity=.4,shape=rectangle,label=$\mathbb{B}_N$,fontscale=0.75]{N}
    \Edge(r+1)(dots)
    \Edge(dots)(i*-1)
    \Edge(i*-1)(i*+1)
    \Edge(i*+1)(dots2)
    \Edge(dots2)(N)
    \Vertex[x=0,y=1.25,shape=rectangle,style={color=white},label=$\sigma_{r+1}$]{sr+1}
    \Vertex[x=2.5,y=1.25,shape=rectangle,style={color=white},label=$\sigma_{i^*-1}$]{si*-1}
    \Vertex[x=3.75,y=1.25,shape=rectangle,style={color=white},label=$\sigma_{i^*+1}$]{si*+1}
    \Vertex[x=6.25,y=1.25,shape=rectangle,style={color=white},label=$\sigma_N$]{sN}
    \Edge(r+1)(sr+1)
    \Edge(i*-1)(si*-1)
    \Edge(i*+1)(si*+1)
    \Edge(N)(sN)
\end{tikzpicture}
\end{equation}
is the state appearing in Eq.~\eqref{psi_final_magic} and it also has bond dimension $\chi$. Notice that site $i^*$ is missing from Eq.~\eqref{psi_out}, and $i^*-1$ is directly linked to $i^*+1$.

Our goal is to optimize the choice of $\ket{\beta_i}$ to maximize the overlap 
\begin{equation}\label{overlap}
    \braket{\Psi_\text{approx}}{\Psi_\text{exact}} = \frac{\left(\bra{+_{i^*}}\otimes\bra{\Tilde{\Phi}^{(N-r-1)}_\text{non-stab}}\right) \hat{O} \ket{\Tilde{\Phi}^{(N-r)}_\text{non-stab}}}{2\mathcal{N}},
\end{equation}
where
\begin{equation}
    \hat{O} = \hat{X}_{i^*}^\frac{1-s}{2}\left(\prod_{i\in\mathcal{S}\setminus\{i^*\}}\hat{\mathrm{CX}}_{i\to i^*}\right) (\hat{S}_{i^*}\daga)^{q^*}\left(\hat{\mathds{1}}+s\prod_{i\in\mathcal{S}}\hat{Z}_i\right).
\end{equation}
We notice that the projector $\hat{\mathds{1}}+s\prod_{i\in\mathcal{S}}\hat{Z}_i$ destroys all computational basis states with parity different from $s$. Then, the sequence of CNOT gate forces the qubit $i^*$ to the state $\ket{s}$. As a consequence, we can rewrite
\begin{equation}
\begin{split}
    \hat{O} &= \hat{X}_{i^*}^b\ket{s_{i^*}}\left(\bra{+_{i^*}}+\bra{-_{i^*}}\right)_{i^*}(\hat{S}_{i^*}\daga)^{q^*}\left(\hat{\mathds{1}}+s\prod_{i\in\mathcal{S}}\hat{Z}_i\right)\\
    &= \ket{+_{i^*}}\left(\bra{+_{i^*}}+i^{-q^*}\bra{-_{i^*}}\right)\left(\hat{\mathds{1}}+s\prod_{i\in\mathcal{S}}\hat{Z}_i\right).
    \end{split}
\end{equation}
The matrix element of Eq.~\eqref{overlap} is now given by the tensor contraction
\begin{multline}
    \braket{\Psi_\text{approx}}{\Psi_\text{exact}} \\
    = \frac{\mathcal{N}}{2}\;
\begin{tikzpicture}[baseline={([yshift=-0.5ex]current bounding box.center)}, scale=0.7]
    \Vertex[x=0,y=2.5,color=blue,opacity=.4,shape=rectangle,label=$\mathbb{B}^*_{r+1}$,fontscale=0.75]{r+1} 
    \Vertex[x=1.25,y=2.5,shape=rectangle,style={color=white},label=$\dots$]{dots}
    \Vertex[x=2.5,y=2.5,color=blue,opacity=.4,shape=rectangle,label=$\mathbb{B}^*_{i^*-1}$,fontscale=0.75]{i*-1}
    \Vertex[x=5,y=2.5,color=blue,opacity=.4,shape=rectangle,label=$\mathbb{B}^*_{i^*+1}$,fontscale=0.75]{i*+1}
    \Vertex[x=6.25,y=2.5,shape=rectangle,style={color=white},label=$\dots$]{dots2}
    \Vertex[x=7.5,y=2.5,color=blue,opacity=.4,shape=rectangle,label=$\mathbb{B}^*_N$,fontscale=0.75]{N}
    \Edge(r+1)(dots)
    \Edge(dots)(i*-1)
    \Edge(i*-1)(i*+1)
    \Edge(i*+1)(dots2)
    \Edge(dots2)(N)
    \Vertex[x=0,y=1.25,shape=rectangle,color=gray,opacity=0.5,label=$\mathbb{W}_{r+1}$,fontscale=0.75]{sr+1}
    \Vertex[x=1.25,y=1.25,shape=rectangle,style={color=white},label=$\dots$]{sdots}
    \Vertex[x=2.5,y=1.25,shape=rectangle,color=gray,opacity=0.5,label=$\mathbb{W}_{i^*-1}$,fontscale=0.75]{si*-1}
    \Vertex[x=3.75,y=1.25,shape=rectangle,color=gray,opacity=0.5,label=$\mathbb{W}_{i^*}$,fontscale=0.75]{si*}
    \Vertex[x=5,y=1.25,shape=rectangle,color=gray,opacity=0.5,label=$\mathbb{W}_{i^*+1}$,fontscale=0.75]{si*+1}
    \Vertex[x=6.25,y=1.25,shape=rectangle,style={color=white},label=$\dots$]{sdots2}
    \Vertex[x=7.5,y=1.25,shape=rectangle,color=gray,opacity=0.5,label=$\mathbb{W}_{N}$,fontscale=0.75]{sN}
    \Edge(r+1)(sr+1)
    \Edge(i*-1)(si*-1)
    \Edge(i*+1)(si*+1)
    \Edge(N)(sN)
    \Edge(sr+1)(sdots)
    \Edge(sdots)(si*-1)
    \Edge(si*-1)(si*)
    \Edge(si*)(si*+1)
    \Edge(si*+1)(sdots2)
    \Edge(sdots2)(sN)
    \Vertex[x=0,color=orange,opacity=.7,shape=rectangle,label=$\mathbb{A}_{r+1}$,fontscale=0.75]{kr+1} 
    \Vertex[x=1.25,shape=rectangle,style={color=white},label=$\dots$]{kdots}
    \Vertex[x=2.5,color=orange,opacity=.7,shape=rectangle,label=$\mathbb{A}_{i^*-1}$,fontscale=0.75]{ki*-1}
    \Vertex[x=3.75,color=orange,opacity=.7,shape=rectangle,label=$\mathbb{A}_{i^*}$,fontscale=0.75]{ki*}
    \Vertex[x=5,color=orange,opacity=.7,shape=rectangle,label=$\mathbb{A}_{i^*+1}$,fontscale=0.75]{ki*+1}
    \Vertex[x=6.25,shape=rectangle,style={color=white},label=$\dots$]{kdots2}
    \Vertex[x=7.5,color=orange,opacity=.7,shape=rectangle,label=$\mathbb{A}_N$,fontscale=0.75]{kN}
    \Edge(kr+1)(kdots)
    \Edge(kdots)(ki*-1)
    \Edge(ki*-1)(ki*)
    \Edge(ki*)(ki*+1)
    \Edge(ki*+1)(kdots2)
    \Edge(kdots2)(kN)
    \Edge(kr+1)(sr+1)
    \Edge(ki*-1)(si*-1)
    \Edge(ki*)(si*)
    \Edge(ki*+1)(si*+1)
    \Edge(kN)(sN)
\end{tikzpicture}
\end{multline}
where we introduced a matrix-product operator (MPO) defined by
\begin{subequations}
\begin{equation}
    \mathbb{W}_{i} =
        \begin{pmatrix}
            \hat{\mathds{1}} & 0\\ 0 & \hat{M}_i
        \end{pmatrix} \quad \text{for} \,\, i\neq i^*
\end{equation},
\begin{equation}
    \mathbb{W}_{i^*} = 
        \begin{pmatrix}
            \bra{+_{i^*}}+\bra{-_{i^*}} & 0\\ 0 & s\left(\bra{+_{i^*}}-\bra{-_{i^*}}\right)
        \end{pmatrix},
\end{equation}
\end{subequations}
where $\hat{M}_i = \hat{Z}$ if $i\in\mathcal{S}$ and $\hat{M}_i = \mathds{1}$ otherwise. In the previous equation, the boundary matrices $\mathbb{W}_{r+1}$ and $\mathbb{W}_N$ are understood to be a row and a column vector arrays respectively to contract correctly.

We set up the variational problem as follows. For each site $j\neq i^*$, we maximize the overlap as a function of $\mathbb{B}_j$ alone fixing all others and imposing the normalization constraint $\braket{\Tilde{\Phi}^{(N-r-1)}_\text{non-stab}}{\Tilde{\Phi}^{(N-r-1)}_\text{non-stab}}=1$. This is easily done with standard methods~\cite{Schollwock_2011} if we assume that $\ket{\Tilde{\Phi}^{(N-r-1)}_\text{non-stab}}$ is put in mixed-canonical form centered around site $j$, in which case the solution of the variational step reads
\begin{equation}\label{variational_step}
\begin{tikzpicture}[baseline={([yshift=-0.5ex]current bounding box.center)}, scale=0.7]
    \Vertex[x=-5,y=1.25,color=blue,opacity=.4,shape=rectangle,label=$\mathbb{B}_j$,fontscale=0.75]{jj}
    \Vertex[x=-3.75,y=1.25,shape=rectangle,style={color=white},label=$\propto$,fontscale=1.5]{r}
    \Vertex[x=-5,y=2.5,shape=rectangle,style={color=white}]{up}
    \node[draw, minimum size=0pt, inner sep=0pt] (A) at (-5.833333333,1.25) {};
    \Edge(A)(jj)
    \Edge(jj)(r)
    \Edge(jj)(up)
    \Vertex[x=-2.5,y=2.5,color=blue,opacity=.4,shape=rectangle,label=$\mathbb{B}^*_{r+1}$,fontscale=0.75]{r+1} 
    \Vertex[x=-1.25,y=2.5,shape=rectangle,style={color=white},label=$\dots$]{dots}
    \Vertex[x=0,y=2.5,color=blue,opacity=.4,shape=rectangle,label=$\mathbb{B}^*_{j-1}$,fontscale=0.75]{i*-1}
    \Vertex[x=1.25,y=2.5,shape=rectangle,style={color=white}]{i}
    \Vertex[x=2.5,y=2.5,color=blue,opacity=.4,shape=rectangle,label=$\mathbb{B}^*_{j+1}$,fontscale=0.75]{i*+1}
    \Vertex[x=3.75,y=2.5,shape=rectangle,style={color=white},label=$\dots$]{dots2}
    \Vertex[x=5,y=2.5,color=blue,opacity=.4,shape=rectangle,label=$\mathbb{B}^*_N$,fontscale=0.75]{N}
    \Edge(r+1)(dots)
    \Edge(dots)(i*-1)
    \Edge(i*-1)(i)
    \Edge(i)(i*+1)
    \Edge(i*+1)(dots2)
    \Edge(dots2)(N)
    \Vertex[x=-2.5,y=1.25,shape=rectangle,color=gray,opacity=0.5,label=$\mathbb{W}_{r+1}$,fontscale=0.75]{sr+1}
    \Vertex[x=-1.25,y=1.25,shape=rectangle,style={color=white},label=$\dots$]{sdots}
    \Vertex[x=0,y=1.25,shape=rectangle,color=gray,opacity=0.5,label=$\mathbb{W}_{j-1}$,fontscale=0.75]{si*-1}
    \Vertex[x=1.25,y=1.25,shape=rectangle,color=gray,opacity=0.5,label=$\mathbb{W}_{j}$,fontscale=0.75]{si*}
    \Vertex[x=2.5,y=1.25,shape=rectangle,color=gray,opacity=0.5,label=$\mathbb{W}_{j+1}$,fontscale=0.75]{si*+1}
    \Vertex[x=3.75,y=1.25,shape=rectangle,style={color=white},label=$\dots$]{sdots2}
    \Vertex[x=5,y=1.25,shape=rectangle,color=gray,opacity=0.5,label=$\mathbb{W}_{N}$,fontscale=0.75]{sN}
    \Edge(r+1)(sr+1)
    \Edge(i*-1)(si*-1)
    \Edge(i*+1)(si*+1)
    \Edge(N)(sN)
    \Edge(sr+1)(sdots)
    \Edge(sdots)(si*-1)
    \Edge(si*-1)(si*)
    \Edge(si*)(si*+1)
    \Edge(si*+1)(sdots2)
    \Edge(sdots2)(sN)
    \Vertex[x=-2.5,color=orange,opacity=.7,shape=rectangle,label=$\mathbb{A}_{r+1}$,fontscale=0.75]{kr+1} 
    \Vertex[x=-1.25,shape=rectangle,style={color=white},label=$\dots$]{kdots}
    \Vertex[x=0,color=orange,opacity=.7,shape=rectangle,label=$\mathbb{A}_{j-1}$,fontscale=0.75]{ki*-1}
    \Vertex[x=1.25,color=orange,opacity=.7,shape=rectangle,label=$\mathbb{A}_{j}$,fontscale=0.75]{ki*}
    \Vertex[x=2.5,color=orange,opacity=.7,shape=rectangle,label=$\mathbb{A}_{j+1}$,fontscale=0.75]{ki*+1}
    \Vertex[x=3.75,shape=rectangle,style={color=white},label=$\dots$]{kdots2}
    \Vertex[x=5,color=orange,opacity=.7,shape=rectangle,label=$\mathbb{A}_N$,fontscale=0.75]{kN}
    \Edge(kr+1)(kdots)
    \Edge(kdots)(ki*-1)
    \Edge(ki*-1)(ki*)
    \Edge(ki*)(ki*+1)
    \Edge(ki*+1)(kdots2)
    \Edge(kdots2)(kN)
    \Edge(kr+1)(sr+1)
    \Edge(ki*-1)(si*-1)
    \Edge(ki*)(si*)
    \Edge(ki*+1)(si*+1)
    \Edge(kN)(sN)
    \Edge(si*)(i)
\end{tikzpicture}\,\, ,
\end{equation}
where $\mathbb{B}_j$ is determined up to a prefactor that is determined by normalizing the MPS. We then repeat the procedure sweeping through all sites a few times until convergence is reached. At this point, we can evaluate the final overlap $\braket{\Psi_\text{approx}}{\Psi_\text{exact}}$ to obtain the fidelity introduced in Eq.~\eqref{fidelity}. Contracting the right-hand side of Eq.~\eqref{variational_step} has a computational cost that scales as $\mathcal{O}((N-r)\chi^3)$. Since we need to repeat this step multiple times for each site, the overall cost of the iterative optimization scales as $\mathcal{O}( (N-r)^2\chi^3)$.

\section{Optimal way to treat $T$ gates}\label{a:T_gates}
The ICCR algorithm can be applied to circuits doped with $T$ gates using the $T$ gadgets introduced in Sec.~\ref{s:algorithm}. The straightforward implementation of this method requires the introduction of $n_T$ ancilla qubits, where $n_T$ is the number of $T$ gates in the circuit, which increases the computational cost of the simulation. We now show a better strategy to proceed that requires only a single ancilla qubit that can be reused for all $T$ gates.

Consider an initial state $\ket{\Psi_S}$ evolved with a Clifford circuit and then with a single $T$ gate. Using the $T$ gadget replacement, we can reformulate the evolution as in Fig.~\eqref{f:algo_scheme}a, where now the initial state $\ket{\Psi} = \ket{\Psi_S}\ket{+_A}$ is the product of the $N$-physical-qubit state $\ket{\Psi_S}$ and the ancilla-qubit state $\ket{+_A}$, and the projector acts on the ancilla. We implement the ICCR algorithm, which approximates the end-circuit state as
\begin{equation}
    \ket{\Psi_f} = \hat{U}' \ket{\Psi'}.
\end{equation}
We can make some exact statements on both $\ket{\Psi'}$ and $\ket{\Psi_f}$. The renormalized initial state $\ket{\Psi'}$ contains at least one stabilizer qubit (i.e., the target qubit $i^*$) in the factorized state $\ket{+}$ as in Fig.~\ref{f:algo_scheme}d, and can thus be written as $\ket{\Psi'} = \ket{\Phi}\ket{+_{i^*}}$. Then, the final state can always be written as $\ket{\Psi_f}=\ket{\Psi_{f,S}}\ket{+_A}$ because the projector disentangles the ancilla from the physical qubits. If we swap the ancilla with the qubit $i^*$ by defining $\ket{\Psi'_S}\ket{+_A} = \hat{\mathrm{SWAP}}_{i^*\leftrightarrow A}\ket{\Psi'}$, where $\hat{\mathrm{SWAP}}$ is the (Clifford) SWAP gate, we have
\begin{equation}
    \ket{\Psi_{f,S}}\ket{+}_A = \hat{U}'\,\hat{\mathrm{SWAP}}_{i^*\leftrightarrow A} (\ket{\Psi'_S}\ket{+_A}) = \hat{U}'' (\ket{\Psi'_S}\ket{+_A}).
\end{equation}
We see that the Clifford unitary $\hat{U}''$ leaves the ancilla in the state $\ket{+}$. This means that there must exist a Clifford operator $\hat{W}_S$ acting only on $S$ such that $\hat{U}''(\ket{\Psi'_S}\ket{+_A}) = (\hat{W}_S \ket{\Psi'_S} )\ket{+_A}$. Notice however that in general $\hat{U}'' \neq \hat{W}_S \hat{\mathds{1}}_A$. We now show how to determine $\hat{W}_S$.

Since $\hat{U}''$ does not change the state $\ket{+}$ of the ancilla for any choice of the system state $\ket{\Psi'_S}$, we know that either $+\hat{Z}_A$ or $-\hat{Z}_A$ must belong to the stabilizer set of $\hat{U}''$, i.e.,
\begin{equation}
    (\hat{U}'')\daga \hat{Z}_A \hat{U}'' = \pm \hat{Z}_A.
\end{equation}
In general, the same will not hold for $\hat{X}_A$: the unitary gate will map it to a complicated string
\begin{equation}
    (\hat{U}'')\daga \hat{X}_A \hat{U}'' = \pm \left(\prod_{i=1}^N \hat{P}_i\right)\hat{P}_A,
\end{equation}
where $\hat{P}_i\in\{\hat{\mathds{1}}_i,\hat{X}_i,\hat{Y}_i,\hat{Z}_i\}$ and $\hat{P}_A\in\{\hat{X}_A,\hat{Y}_A\}$. Notice that $\hat{P}_A$ cannot be neither $\hat{\mathds{1}}_A$ nor $\hat{Z}_A$ because $\comm{(\hat{U}'')\daga \hat{Z}_A \hat{U}''}{(\hat{U}'')\daga \hat{X}_A \hat{U}''} = (\hat{U}'')\daga \comm{\hat{Z}_A}{\hat{X}_A}\hat{U}'' \neq 0$. Our goal is to find a new Clifford unitary $\hat{\Tilde{U}}$ such that it acts in the same way as $\hat{U}''$, meaning that
\begin{equation}\label{U_tilde}
    \hat{\Tilde{U}}(\ket{\Psi'_S}\ket{+_A})=\hat{U}''(\ket{\Psi'_S}\ket{+_A}),
\end{equation}
but at the same time has $\pm\hat{X}_A$ in its stabilizer set. For each $i$, we define
\begin{equation}
    \hat{C}_i = \begin{cases}
        \hat{\mathds{1}} \qquad\text{if}\, \, \hat{P}_i=\hat{\mathds{1}}_i,\\
        \hat{CX}_{A\to i} \qquad\text{if}\, \, \hat{P}_i=\hat{X}_i,\\
        \hat{CY}_{A\to i} \qquad\text{if}\, \, \hat{P}_i=\hat{Y}_i,\\
        \hat{CZ}_{A\to i} \qquad\text{if}\, \, \hat{P}_i=\hat{Z}_i,\\
    \end{cases}
\end{equation}
where $\hat{CY}_{A\to i}$ and $\hat{CZ}_{A\to i}$ are respectively the controlled $Y$ and $Z$ gates that use the ancilla as reference and $i$ as target. We then introduce
\begin{equation}
    \hat{\Tilde{U}} = \hat{U}''\prod_{i=1}^N \hat{C}_i,
\end{equation}
which clearly satisfies Eq.~\eqref{U_tilde} due to the definition of $\hat{C}_i$. Moreover, it is easily checked that
\begin{equation}\label{stab_Z}
    \hat{\Tilde{U}}\daga \hat{Z}_A\hat{\Tilde{U}} = \pm \hat{Z}_A
\end{equation}
and
\begin{equation}\label{stab_X}
    \hat{\Tilde{U}}\daga \hat{X}_A\hat{\Tilde{U}} = \pm \hat{P}_A.
\end{equation}
Notice that the sign $\pm$ of Eq.~\eqref{stab_X} is not necessarily the same as that of Eq.~\eqref{stab_Z}. Equations~\eqref{stab_Z} and~\eqref{stab_X} imply that the gate $\hat{\Tilde{U}}$ is unable to entangle the system with the ancilla for all initial states. We thus conclude that $\hat{\Tilde{U}} = \hat{W}_S \hat{Q}_A$, where $\hat{Q}_A$ is a single-qubit gate that acts as the identity on $\ket{+_A}$. Finally, the stabilizer tableau of $\hat{W}_S$ is obtained by dropping the rows and columns of the tableau of $\hat{\Tilde{U}}$ corresponding to the ancilla qubit, yielding the final identity
\begin{equation}
\ket{\Psi_{f,S}} = \hat{W}_S\ket{\Psi'_S}
\end{equation}
for the physical qubits alone.

\bibliography{bibliography}

\end{document}